\documentclass{article}

\usepackage{amssymb,amsfonts,amsmath}
\usepackage{cite,enumerate,float,indentfirst}
\usepackage{color}

\def\be{\begin{eqnarray}}
\def\ee{\end{eqnarray}}
\def\nn{\nonumber}

\def\tr{{\rm tr}\,}
\def\Tr{{\rm Tr}\,}

\definecolor{red}{rgb}{1,0,0}
\definecolor{orange}{rgb}{1,0.5,0}
\definecolor{violet}{rgb}{0.7,0,1}

\hoffset= - 1.0in         

\begin{document}

\title{On $(q,t)$-deformation of Gaussian  matrix model}
\author{{\bf A.Morozov$^{a,b,}$}\footnote{morozov@itep.ru}, {\bf A. Popolitov$^{a,b,c}$}\footnote{popolit@gmail.com}  \ and  {\bf Sh. Shakirov$^{b,d,e}$}\thanks{shakirov@fas.harvard.edu}}
\date{ }

\maketitle

\vspace{-6.0cm}

\begin{center}
\hfill IITP/TH-05/18\\
\hfill ITEP/TH-07/18
\end{center}

\vspace{4.2cm}

\begin{center}
$^a$ {\small {\it ITEP, Moscow 117218, Russia}}\\
$^b$ {\small {\it Institute for Information Transmission Problems, Moscow 127994, Russia}}\\
$^c$ {\small {\it Department of Physics and Astronomy, Uppsala University, Uppsala, Sweden}}\\
$^d$ {\small {\it Society of Fellows, Harvard University, Cambridge, MA 02138, USA}}\\
$^e$ {\small {\it Mathematical Sciences Research Institute, Berkeley, CA 94720, USA}}
\end{center}

\centerline{ABSTRACT}

\bigskip

{\footnotesize
The recently discovered general formulas for perturbative correlators
in basic matrix models
can be interpreted as the Schur-preservation property of Gaussian measures.
Then substitution of Schur by, say, Macdonald polynomials,
{\it defines} a $q,t$-deformation of the matrix model.
Eigenvalue integral representations and Virasoro-like constraints
are immediate consequences.
}

\bigskip

\bigskip

As a byproduct of recent advances in tensor model calculus \cite{tenmod}
a new kind of realization  was found \cite{exactmamo}
for   polynomial
correlation functions in ordinary matrix models.
The clear advantage is that they directly imply both integrability and
Virasoro-like constraints \cite{UFN3} --
which are usually appearing as two {\it complementary} and {\it badly compatible}
properties, with matrix models lying precisely at the  intersection.
New formulas explicitly describe what this intersection is.

\bigskip

Eigenvalue matrix models  are usually defined
as Gaussian averages over matrices $M$ of
exponentials of traces:
\be
Z\{p\} = \left< \exp \left(\sum_k \frac{p_k P_k^{(M)}}{k} \right)\right> =
\sum_R \Big<\chi_R\{P^{(M)}\}\Big>\cdot \chi_R\{p\}
\label{Zdef}
\ee
where  formula at the r.h.s. is the Cauchy identity for
the sum of Schur functions $\chi_R$ over all Young diagrams $R$.
The time-variables $P^{(M)}_k = \Tr M^k$ are made from the matrix-valued
Miwa variable $M$,  and average  is taken over $M$ with  Gaussian measure.
The two most important examples are the rectangular model of \cite{rect},
where $M$ is the complex-valued $N_1\times N_2$ matrix and
\be
\frac{1}{\mu^{N_1N_2}}\cdot \Big<\ \ldots \Big>_{N_1\times N_2}
=  \int \ \ldots \ e^{-\mu P_1^{(MM^\dagger)}}d^2M
=  \int \ \ldots \ e^{-\mu\, \Tr MM^\dagger }d^2M
\label{rectmo}
\ee
and Hermitian model with $N\times N$ matrix $M=M^\dagger$ and
\be
\frac{1}{\mu^{N^2/2}} \cdot \Big<\ \ldots \Big>_{N}
= \int \ \ldots \ e^{-\frac{\mu}{2} P_2^{(M)}}dM
= \int \ \ldots \ e^{-\frac{\mu}{2}\, \Tr M^2 }dM
\label{hermmo}
\ee

\bigskip

The main claim of \cite{exactmamo}  is that such averages
of Schur functions are again made from the Schur functions:
 \be
\Big<\chi_R\left\{P^{(MM^\dagger)}\right\}\Big>_{N_1\times N_2}
= \frac{1}{\mu^{|R|}}\cdot\frac{D_R(N_1)D_R(N_2)}{d_R}
= \frac{1}{\mu^{|R|}}\cdot
\frac{\chi_R\{p^{(N_1)}\}\cdot \chi_R\{p^{(N_2)}\}}{\chi_R\{\delta_{k,1}\}}
\label{rectcor}
\ee
\be
\Big<\chi_R\left\{P^{(M)}\right\}\Big>_{N}
= \frac{1}{\mu^{|R|/2}}\cdot \frac{D_R(N)\cdot \varphi_R(2^{|R|/2})}{d_R}
= \frac{1}{\mu^{|R|/2}}\cdot\frac{ \chi_R\{\delta_{k,2}\}}{\chi_R\{\delta_{k,1}\}}
\cdot \chi_R\{p^{(N)}\}
\label{hermcor}
\ee
for rectangular and Hermitian model, respectively; we remind that dimensions $D_R$ are particular values of Schur functions $\chi_R$ at important special points
\be
p_k = p_k^{(N)} = N
\ee
which correspond to traces of powers of the identity matrix, $\tr I^k = N$. One can see that, structurally,
\be
\boxed{
\Big< \chi_R\{P^{(M)}\}\Big> \sim \chi_R\{p^{(N)}\}
}
\label{ShuShu}
\ee
where on the l.h.s. we average (integrate) over the matrix-valued "field" $M$,
while on the r.h.s. $N$ is just the matrix size.
Eq.(\ref{ShuShu}) means {\bf the main feature of Gaussian matrix measures is that they preserve Schur functions} in the above sence,
while localizing the integral over Miwa time-variables $M$ to the very special locus $p^{(N)}$.

The proportionality coefficients are built from the values of Schur functions at two other important points:
one is $\delta_{k,1}$,  whose significance is due to the fact that the partition functions
 are easily evaluated at it by the Cauchy rule:
\be
Z_{N_1\times N_2}\{p\}
= \sum_R \frac{1}{\mu^{|R|}}\cdot
\frac{\chi_R\{p^{(N_1)}\}\,\chi_R\{p^{(N_2)}\}} {\chi_R\{\delta_{k,1}\}}
\cdot\chi_R\{p\}\ \longrightarrow
\nn \\
\ \stackrel{p_k = \alpha \delta_{k,1}}{\longrightarrow} \
\sum_R \left(\frac{\alpha}{\mu}\right)^{|R|}\!\!
\chi_R\{p^{(N_1)}\}\,\chi_R\{p^{(N_2)}\} =
\exp \left(\sum_k \frac{\alpha^k}{\mu^k}\cdot \frac{ p_k^{(N_1)}p_k^{(N_2)}}{k}\right) =
\left(1-\frac{\alpha}{\mu}\right)^{-N_1N_2}
\label{rectval}
\ee
\be
Z_{N }\{p\}
= \sum_R \frac{1}{\mu^{|R|/2}}\cdot
\frac{\chi_R\{\delta_{k,2}\}\,\chi_R\{p^{(N)}\}\,\chi_R\{p\}} {\chi_R\{\delta_{k,1}\}}
\ \longrightarrow
\nn \\
\ \stackrel{p_k = \alpha \delta_{k,1}}{\longrightarrow} \
\sum_R \left(\frac{\alpha}{\sqrt{\mu}}\right)^{|R|}\!\!
\chi_R\{\delta_{k,2}\}\,\chi_R\{p^{(N )}\} =
\exp \left(\sum_k \frac{\alpha^k}{\mu^{k/2}}\cdot \frac{\delta_{k,2}\,p_k^{(N)}}{k}\right) =
\exp \left(\frac{\alpha^2N}{2\mu}\right)
\label{hermval}
\ee
Another is $\delta_{k,2}$ which is exactly what needed to allow the absorbtion
of $p_1$ into the shift of the Gaussian distribution in Hermitian model.
These expressions are exactly what would follow from integral realizations (\ref{rectmo}) and
(\ref{hermmo}) respectively. The advantage of (\ref{rectval}) and (\ref{hermval}), however, is that they
are pure algebraic and do not refer to any integrals.

\bigskip

One of the many advantages of spectacular relation (\ref{ShuShu}) is that
it provides a property of Gaussian measures,
which calls for obvious generalizations and can be easily deformed.
For example, one can notice that the distinguished locus
\be
p_k = p_k^{(N)} = N
\ee
is a particular $(t=1)$ case of the topological locus
\be
p_k=\pi^{(N)}_k = \frac{[Nk]}{[k]} = \frac{t^{-Nk}-t^{Nk}}{t^{-k}-t^{k}}
\ee
which plays a big role in group theory and its applications,
like Chern-Simons and knot theories, where it appears as the argument of a wide variety of functions generalizing the Schur polynomials. Thus it is a natural suggestion to deform the Gaussian measure by first changing $p^{(N)}$ at the r.h.s. of (\ref{ShuShu}) to $\pi^{(N)}$
and then replacing the Schur functions at the l.h.s. by another basis -- well-known such as Hall-Littlewood, Jack, Macdonald, or more general such as Askey-Wilson \cite{askeywilson},
elliptic \cite{ellipticMc}, affine \cite{affineMc},
generalized \cite{genmac} or triple \cite{3Mc} Macdonald etc.
This can be considered as a {\bf definition} of  $t$-, $\beta$-, $(q,t)$-
and further deformed Gaussian measures.

\bigskip

Clearly, these measures implicitly appear in many modern problems --
and their explicit definition implies new conjectures about various fields
of theoretical physics and identify what is in common between them. This is especially important because so defined $(q,t)$-deformed matrix models
preserve most important features of the ordinary ones.
They possess "eigenvalue" integral representations
(in terms of Jackson integrals/sums)
and satisfy difference equations, which are straightforward deformations
of the differential Virasoro constraints.
Seemingly lost are only integrability properties --
deformed partition functions are not just the ordinary
KP/Toda $\tau$-functions,
but this only emphasizes the need for "quantum $\tau$-functions",
which were suggested long ago \cite{quanttau} but remain under-investigated.

\bigskip

In the $(q,t)$-case we substitute (\ref{Zdef}) by
\be
\boxed{
{\cal Z}\{p\} = \sum_R \frac{1}{||{\cal M}_R||^2}\cdot \Big<{\cal M}_R\{P^{M}\} \Big>\cdot {\cal M}_R\{p\}
}
\label{calZdef}
\ee
where
\be
||{\cal M}_R||^2 = \prod\limits_{(i,j) \in R} \dfrac{t^{-R^T_j+i}q^{-R_i+j-1} - t^{R^T_j-i}q^{R_i-j+1}}{t^{-R^T_j+i-1}q^{-R_i+j} - t^{R^T_j-i+1}q^{R_i-j}}
\ee
and Cauchy identity becomes
\be
\sum_R \frac{x^{|R|}}{||{\cal M}_R||^2}\cdot {\cal M}_R\{p\} {\cal M}_R\{p^{\prime}\} = \exp\left( \sum\limits_{k = 1}^{\infty} \ \dfrac{t^{-k} - t^{k}}{q^{-k} - q^{k}} \ \dfrac{p_k p^{\prime}_k}{k} \ x^k \right)
\ee
The central point is the definition of averages by the most naive deformation of
(\ref{ShuShu}):
\be
\boxed{
\Big<{\cal M}_R\left\{P^{(MM^\dagger)}\right\}\Big>_{N_1\times N_2}
= \frac{1}{\mu^{|R|}}\cdot
\frac{{\cal M}_R\{\pi^{(N_1)}\}\cdot {\cal M}_R\{\pi^{(N_2)}\}}{{\cal M}_R\{\delta^{*}_{k,1}\}}
}
\label{qtrectmo}
\ee
for rectangular model, and
\be
\boxed{
\Big<{\cal M}_R\left\{P^{(M)}\right\}\Big>_{N}
= \frac{1}{\mu^{|R|/2}}\cdot\frac{ {\cal M}_R\{\delta^{*}_{k,2}\}}{{\cal M}_R\{\delta^{*}_{k,1}\}}
\cdot {\cal M}_R\{\pi^{(N)}\}
}
\label{qthermmo}
\ee
for Hermitian one. The points
\be
\boxed{
\delta^{*}_{k,n} = n\cdot \dfrac{(q^{-1} - q)^{k/n}}{t^{-k} - t^{k}}\cdot \delta_{k|n}
}
\label{qttimes}
\ee
where $\delta_{k|n}$ selects $k$ divisible by $n$,
are the appropriate deformations of $\delta_{k,1}$ and $\delta_{k,2}$.
They are distinguished by preservation of the
factorized expressions for ${\cal M}_R\{\delta^{*}_{k,1}\}$
and straightforward deformation of (\ref{rectval}) and (\ref{hermval}):

\be
Z_{N_1\times N_2}\{p\} \ \stackrel{p_k = \,\alpha\, \delta^{*}_{k,1}}{\longrightarrow} \
\sum_R \left(\frac{\alpha}{\mu}\right)^{|R|}\!\! \ \dfrac{1}{||{\cal M}_R||^2} \
{\cal M}_R\{\pi^{(N_1)}\}\,{\cal M}_R\{\pi^{(N_2)}\} = \nn \\
= \exp \left(\sum_k \frac{\alpha^k}{\mu^k}\cdot \dfrac{t^{-k}-t^k}{q^{-k}-q^k} \ \frac{ \pi_k^{(N_1)}\pi_k^{(N_2)}}{k}\right) =
\prod\limits_{m = 0}^{\beta-1}\prod\limits_{i=1}^{N_1}\prod\limits_{j=1}^{N_2} \left( 1 - \frac{\alpha}{\mu} \ q^{2m+1} t^{2i+2j-N_1-N_2+1} \right)^{-1}
\ee
and
\be
Z_{N }\{p\} \ \stackrel{p_k = \,\alpha\, \delta^{*}_{k,1}}{\longrightarrow} \
\sum_R \left(\frac{\alpha}{\sqrt{\mu}}\right)^{|R|}\!\! \ \dfrac{1}{||{\cal M}_R||^2} \
{\cal M}_R\{\delta^*_{k,2}\}\,{\cal M}_R\{\pi^{(N )}\} =
\exp \left(\ \sum_{{\rm even}\ k}\ \frac{\alpha^k}{\mu^{k/2}}\cdot \dfrac{t^{-k}-t^k}{q^{-k}-q^k} \ \frac{\pi_k^{(N)}\!\!\cdot  \delta^*_{k,2} }{k}\right) =
\!\!\!\!\!\!\!\!\!\!\!\nn \\ =
\exp \left(-\sum_{k=1}^\infty \frac{\Big(-\frac{\alpha^2(q-q^{-1})}{ \mu }\Big)^k}{k(q^{2k}-q^{-2k})}
\cdot   \dfrac{t^{2Nk}-t^{-2Nk}}{t^{2k}-t^{-2k}} \right)
 =\prod\limits_{i=1}^{N} \exp_{q^2} \left(\frac{\alpha^2\,t^{4i-2N-2  }}{[2]_q\cdot\mu}  \right)
 \ \ \ \ \ \ \ \ \
\ee
where

$$
e_q\left(\frac{x}{q-q^{-1}}\right) = \sum_{n=0}^\infty \frac{x^n\,q^{-n(n-1)/2}}{(q^n-q^{-n})!}
= \prod_{n=0}^\infty (1+q^{2n+1}x)^{^{-1}}
$$
\smallskip\\
Perhaps even more convincingly, with this choice of $\delta^{*}$, the model still enjoys an integral representation -- a $(q,t)$ analogue of (\ref{hermmo}), in eigenvalue form -- for the correlators:
\be
\Big<{\cal M}_R\left\{P^{(M)}\right\}\Big>_{N}
= \dfrac{1}{Z} \ \left( \prod\limits_{i = 1}^{N} \int\limits_{-\nu}^{\nu} \rho(x_i) \ d_q x_i \right) \ {\cal M}_R\left\{p_k = x_1^k + \ldots + x_N^k \right\} \ \ \prod\limits_{m = 0}^{\beta-1} \prod\limits_{i \neq j} \ (x_i - q^m x_j)
\label{integralform}
\ee
where $t = q^{\beta}$, $\nu = (1-q)^{\frac{-1}{2}}$, function $\rho(x)$ is the $q$-deformed Gaussian (normal) distribution,
\begin{align}
\rho(x) = \prod\limits_{m = 0}^{\infty} (1 - q^{2m + 2} x^2 \nu^{-2} )
\label{qnormal}
\end{align}
$Z$ is the normalization constant -- partition function of the model:
\be
Z
= \left( \prod\limits_{i = 1}^{N} \int\limits_{-\nu}^{\nu} \rho(x_i) \ d_q x_i \right) \ \ \prod\limits_{m = 0}^{\beta-1} \prod\limits_{i \neq j} \ (x_i - q^m x_j)
\label{integralform}
\ee
and the $q$-deformed (Jackson) symmetric integral is defined by
\begin{align*}
\int\limits_{-\nu}^{\nu} f(x) \ d_q x \ = \ (1-q) \sum\limits_{n = 0}^{\infty} \ \nu q^n \ \big[ f(\nu q^n) + f( - \nu q^n) \big]
\end{align*}
Note that $\nu \rightarrow \infty$ and $\rho(x) \rightarrow e^{-x^2/2}$ as $q \rightarrow 1$. One can also $(q,t)$-deform other properties of the Hermitian matrix model, such as linear Virasoro constraints and bilinear Toda equations; this will be done elsewhere.

\bigskip

To conclude, we reformulated Gaussian matrix models
as Schur-preserving measures in the sense of (\ref{ShuShu})
and suggested to define their deformations as
Macdonald-preserving ones.
Exact definition of the typical $(q,t)$-model is provided by
(\ref{calZdef}) and (\ref{qtrectmo})-(\ref{qttimes}).
Detailed description of the technical properties of such
deformations
and their comparison with more traditional definitions {\it a la} \cite{qtmods}
will be presented elsewhere.
These deformed models are already very important in {\it some} applications,
like conformal theories and associated Dotsenko-Fateev (conformal) matrix models,
in AdS and AGT dualities, in network models, in representation theory of
DIM algebras etc --
but the purpose of this letter is to introduce the general notion,
which we expect to have a very broad value,
without reference to any particular applications. It is straightforward to derive analogous formulas in other above-mentioned
deformations of Gaussian measures -- more modest and more general than the
Macdonald-based $(q,t)$-one. Of separate interest can be straightforward generalization to Aristotelian tensor models.

\section*{Acknowledgements}

This work was supported by the Russian Science Foundation (grant No.16-12-10344).

\end{document}